\documentclass[twocolumn,showpacs,amssymb,amsmath,nobibnotes,prl]{revtex4}

\usepackage{graphicx}
\usepackage{xcolor}
\usepackage{bm}%

\begin{document}
\title{The Magneto-Rotational Decay Instability in Keplerian Disks}
\author{Yuri Shtemler} \email{yshtemler@gmail.com}
\author{Edward Liverts} \email{eliverts@bgu.ac.il} \author{Michael Mond} \email{mond@bgu.ac.il}\affiliation{Department of Mechanical
Engineering, Ben-Gurion University of the Negev, Beer-Sheva 84105,
Israel}
\date{\today}

\begin{abstract}
The saturation of the magnetorotational (MRI) instability in thin Keplerian disks through three-wave resonant interactions is introduced and discussed. 
That mechanism is a natural generalization of the fundamental decay instability discovered five decades ago for infinite, homogeneous and immovable plasmas. 
The decay instability relies on the energy transfer from the MRI to stable slow Alfv\'{e}n-Coriolis (AC) as well as magnetosonic (MS) waves. A second order forced Duffing amplitude equation for the initially unstable MRI as well as two first order equations for the other two waves are derived. The solutions of those equations exhibit bounded bursty nonlinear oscillations for the MRI as well as unbounded growth for the linearly stable slow AC and MS perturbations, thus giving rise to the magneto-rotational decay instability (MRDI). 

\end{abstract}

\maketitle

\textit{Introduction - }
The magnetorotational instability (MRI) {\cite{Velikhov}-\cite{Chandra} is believed to play an important role in the dynamical evolution of thin astrophysical disks {\cite{Balbus1}-\cite{Balbus3}. The analytical understanding of the processes that are responsible for the nonlinear evolution of the MRI is therefore crucial for assessing the true importance of that linear instability to such phenomena as turbulence generation in the disk and angular momentum transfer. First attempts to analyze the nonlinear evolution of the MRI focused on the dissipative saturation of the instability (\cite{KnoblochJulien}- \cite{Umurhan et al}) in environments that are characteristic of laboratory experiments. Recently however, a non dissipative mechanism has been proposed in the context of a thin disk geometry, according to which the MRI saturates to bounded bursty non linear oscillations by non resonantly driving a zero frequency magnetosonic (MS) wave  (\cite{7}-\cite{Livertsetal}).  The scope of the non dissipative mechanism of interacting waves is widened in the current work to include resonant interactions of three linear eigen-oscillations of the system. Extending thus the weakly nonlinear analysis entails a surprising result. While the amplitude of the original MRI saturates via periodical nonlinear oscillations just as in the non-resonant case, it is shown in the current work that the amplitudes of the other two \textit{linearly stable} modes that participate in the resonant triad may grow exponentially through the \textit{nonlinear} magneto-rotational decay instability (MRDI) mechanism. This result provides a natural generalization of the decay instability mechanism discovered five decades ago for infinite, homogeneous, and immovable plasmas \cite{Galeev}, to the geometry of thin, rotating, and axially stratified disks. The resonantly interacting triads of eigenmodes may serve therefore as building blocks of a turbulence model in thin magnetized disks.

\textit{The Physical Model - } The thin disk asymptotic expansion procedure \cite{1}-\cite{5} is applied to the magnetohydrodynamic (MHD) equations in order to study the weakly nonlinear evolution of the MRI in Keplerian disks that are subject to the action of an axial magnetic field. A detailed description of that procedure and its results for the steady-state as well as the linear problem is presented in \cite{6}. The main results are hereby summarized:
1. Steady-State: Assuming axially isothermal steady-state the normalized mass density profiles are given by $n(r,\zeta)=N(r)\Sigma(\eta)$, where $\Sigma(\eta)=e^{-\eta ^2/2}$, $N(r)$ is an arbitrary function of $r$, the radial coordinate, $\eta =\zeta/H(r)$, $\zeta=z/\epsilon$ is the stretched axial coordinate, and $H(r)$ is the semi thickness of the disk. The latter [or alternatively the temperature profile $T(r)$] is an arbitrary function of $r$.
2. Linear perturbations: Modifying the axial mass density profile to ${\bar\Sigma} (\eta) =sech^2\eta$ enables the analytical solution of the linearized set of equations for small perturbations. The resulting eigenmodes are thus divided into two families. The first family, the Alfv\'{e}n-Coriolis (AC) one, represents in-plain perturbations and includes two sets of axially discrete modes. The fast AC modes are stable while the slow AC modes may become unstable. The number of unstable slow AC modes is determined by the local plasma beta which is given by $\beta(r)=\beta_0 N(r)C_s^2(r)/B_z^2(r)$ where $\beta_0$ is the beta value at some reference radius, and $C_s(r)$ and $B_z(r)$ are some arbitrary profiles of the sound velocity and the axial steady-state magnetic field, respectively. Thus, the threshold for exciting $k$ unstable modes is given by $\beta _{cr}^k=k(k+1)/3, k=1,2,\ldots$. It is those unstable slow AC modes that constitute the MRI. The eigenfunctions of both sets of AC modes may be expressed in terms of the Legendre polynomials. Of particle importance is the fact that for $\beta=\beta _{cr}^k$ the k-th eignevalue of the slow AC modes is zero with multiplicity two.  The other family of eigen-oscillations in thin Keplerian disks includes the vertical magnetosonic (MS) modes. The latters are stable, possess a continuous spectrum, and their eigenfunctions are localized about the mid-plain and may be expressed in terms of some Hypergeometric functions. The two families of the linear eigenmodes, namely the AC and the MS modes, are the building blocks of the nonlinear analysis to be unfolded in the next sections.

\textit{Resonant interactions - } The scenario that is introduced in the current work is the following: a large amplitude MRI forms a triad of resonantly interacting modes with a stable fast or slow AC mode, and a stable MS wave. Such interaction is a direct result of the influence of the perturbed in-plain magnetic pressure gradients on the acoustic modes, and the simultaneous axial convection of the AC modes by the acoustic perturbations. Such mechanism underlies the well-known decay instability in plasmas that has been discovered five decades ago \cite{Galeev} and was shown to be of a fundamental nature. Thus, to illustrate the main idea, following \cite{Sagdeev}  consider a parent Alfv\'{e}n wave with amplitude $a_{1}(t)$, and two daughter waves, one of which is another Alfv\'{e}n wave with amplitude $a_{2}(t)$, while the other one is a sound wave with amplitude $a_s(t)$, all co-exist in an infinite uniform, and immovable plasma. A resonant interaction between those three modes occurs if the following resonant conditions are satisfied: $\omega _{2}=\omega _{1}+\omega _s$, and $k_{2}=k_{1}+k_s$. Thus, assuming that the amplitudes vary on a slower time scale than each of the inverse eigen-frequencies, the equations that describe the evolution of the interacting triad may be cast in the following way \cite{Sagdeev}:
\begin{eqnarray}
\frac{da_{1}}{d\tau}&=&i\Gamma a_{2}a_s^* \label{parent}\\
\frac{da_{2}}{d\tau}&=&-i\Gamma a_sa_{1}\label{duaghter1}\\
\frac{da_{s}}{d\tau}&=&i\Gamma a_{1}^*a_{2},\label{daughter2}
\end{eqnarray}
where $\tau$ is a slow time variable. The solution of the above set under initial conditions that $a_{1}$ is much bigger than the other two amplitudes is characterized by cycles of exponential growth of $a_{2}$ and $a_s$ and decay of $a_{1}$, followed by the saturation and decay of the formers and restitution of the latter. During those portions of the cycles that are marked by exponential growth of the daughter waves, $a_{2}$ and $a_s$ grow as $e^{\nu \tau}$ where $\nu =\Gamma a_{1}(\tau=0)$.

Back to thin rotating disks and the MRI, the physics of resonantly interacting triads of eigenmodes is in principle similar to that described above. For simplicity it is assumed that the $\beta$ value of the system is just above the first threshold for instability. Consequently, there is just one unstable MRI mode, characterized by axial wave number $k=1$. The role of the large amplitude parent mode is played therefore by the $k=1$ MRI, while the daughter waves are stable $k=2$ slow AC and MS modes. Thus, contributions of the various modes to the perturbations may be expressed in the following way:
\begin{eqnarray}
\delta B_{\perp}(z,t)=f_1(\zeta,\tau)+f_2(\zeta,\tau)e^{-{\emph i}\omega _2 t}
\label{magnetic}\\
\delta \rho (z,t)=g_1(\zeta,\tau)+g_2(\zeta,\tau)e^{-{\emph i}\omega _2 t}
\label{velocity}
\end{eqnarray}
Equations (\ref{magnetic}) and (\ref{velocity}) describe the AC and MS modes, respectively. The first term on the right hand side of eq. (\ref{magnetic}) represents the parent MRI mode, whose real part of the frequency is zero ($\omega _1 =0$), while the second term describes the contribution of the $k=2$ daughter slow AC mode that is characterized by the eigenvalue $\omega_2$. A main difference from the classical infinite plasma case is the presence of the first term on the right hand side of eq. \ref{velocity} that represents the zero-frequency MS perturbations that are inevitably non resonantly driven by the parent MRI (see \cite{Livertsetal}). The second term describes the contribution of the MS eigenmode with frequency $\omega_2$ so that the resonant condition on the frequencies is fulfilled due to the continuous nature of the MS spectrum. Time $t$ is normalized with the local inverse rotation frequency of the disk $\Omega ^{-1}(r)$, and the slow time is defined as $\tau =\gamma t$ where $\gamma <<1$ is the growth rate of the parent MRI normalized with $\Omega(r)$.

The amplitudes of the various modes in eqs. (\ref{magnetic}) and (\ref{velocity}) may be postulated to be of the following form:
\begin{eqnarray}
f_1(\zeta,\tau)&=&a_1(\tau)P_1(\zeta)+a_2^*(\tau)a_s(\tau)\psi_{2,s}(\zeta)\nonumber\\
& &\mbox{}+a_1^3(\tau)\psi_{1,1}(\zeta)\label{mri}\\
f_2(\zeta,\tau)&=&a_2(\tau)P_2(\zeta)+a_1(\tau)a_s(\tau)\psi_{1,s}(\zeta)\label{acmode}\\
g_2(\zeta,\tau)&=&a_s(\tau)Q_2(\zeta)+a_1(\tau)a_2(\tau)\psi_{1,2}(\zeta)\label{msmode}\\
g_1(\zeta,\tau)&=&a_1^2(\tau)\phi_{1,1}(\zeta)\label{msmode0}
\end{eqnarray} 
The first terms on the right hand sides of eqs. (\ref{mri})-(\ref{msmode}) represent the three linear modes that participate in the resonantly interacting triad where $P_i, i=1,2$ are the eigenfunctions of the MRI and the stable slow AC mode (both expressed, as mentioned above, in terms of the Legendre polynomials), while $Q_2$ is the eigenfunction of the daughter MS mode (expressed in terms of hypergeometric functions). The second terms on the right hand side of (\ref{mri})-(\ref{msmode}) describe the nonlinear resonant interactions through the yet unknown coupling functions $\psi_{i,j}(\zeta), i,j=1,2,s$. Equation (\ref{msmode0}) describes the zero-frequency MS wave that is non resonantly forced by the parent MRI, while the last term on the right hand side of eq. (\ref{mri}) describe its back reaction on the MRI. It should finally be emphasized that unlike in the classical decay instability, since the parent MRI is of zero frequency, $a_1(\tau)$ be assumed to be real. The other two amplitudes are generally complex-valued.

During the linear stage all three modes are independent of each other so that $a_1(\tau)=a_{1}^+e^{\tau}+a_{1}^{-}e^{-\tau}$ (this form of $a_1$ echoes the multiplicity 2 of the corresponding eigenvalue for $\gamma =0$), while $a_2$ and $a_s$ are constants. However, as $a_1$ grows, the nonlinear terms become progressively more important and the temporal behavior of the amplitudes change significantly. It is thus the main goal of the current work to derive the equations that govern the dynamical evolution of the three amplitudes $a_1(\tau)$, $a_2(\tau)$, and $a_s(\tau)$.

Guided by the equations of the classical decay instability [i.e. eqs. (\ref{parent})-(\ref{daughter2})], the equations for $a_2(t)$ and $a_s(t)$ are postulated to be of the following form:
\begin{eqnarray}
\frac{da_{2}}{d\tau}&=&-i\Gamma_2 a_sa_{1}\label{daughter11}\\
\frac{da_{s}}{d\tau}&=&i\Gamma_s a_{1}a_{2}.\label{daughter21}
\end{eqnarray}
The equation for $a_1(\tau)$ however is different from its classical counterpart. First, $a_1$ is the amplitude of the MRI mode slightly above the instability threshold where, as indicated above, the eigenvalue is zero with multiplicity two. Hence, the equation for $a_1$ is expected to be of second order (\cite{Livertsetal}, \cite{9}, \cite{Stefani}). Furthermore, that equation has to include the influence of the driven zero-frequency magnetosonic perturbations. Taking all that into acount, and recalling the $a_1$ is real, the equation for $a_1$ is:
\begin{equation}
\frac{d^2a_1}{d\tau ^2}= a_1 +E a_1^3 +\Gamma _1 (a_2a_s^*+a_2^*a_s).
\label{parent11}
\end{equation}
The first term on the right hand side of the last equation describes the two linear modes (one exponentially growing, the MRI, and the other one evanescent) that coalesce at the threshold to a double zero eigenvalue. The second term describes the contribution of the driven zero-frequency MS perturbations, while the last two terms mark the resonant interaction with the other two modes of the triad. 

The calculation of the four real coupling coefficients in eqs. (\ref{daughter11})-(\ref{parent11}), namely $\Gamma_1,\Gamma_2,\Gamma_s$ and $E$, starts by realizing that those equations are written by tacitly assuming some ordering scheme among the various amplitudes. Thus, recalling that 
$\tau = \gamma t$, all terms in eqs. (\ref{daughter11})-(\ref{parent11}) are of the same order if the amplitude of the parent MRI is proportional to $\gamma  $ while the corresponding amplitudes of the daughter modes are proportional to $\gamma ^{3/2}$ and $\gamma ^{3/2}$. Equations (\ref{daughter11})-(\ref{parent11}) are inserted now into the reduced thin-disk MHD equations \cite{6} which are subsequently solved order by order in  $\gamma$. Not surprising, the lowest order reproduces the linear results. The next order yields four non homogeneous ordinary differential equations for the coupling functions $\psi_{1,2}(\zeta),\psi_{1,s}(\zeta),\psi_{2,s}(\zeta)$ and $\psi_{1,1}(\zeta)$.
The four solvability conditions for those equations (that provide a generalization of the resonant condition on the wave vectors in the classical case), result in four values for the coupling coefficients. 
As expected,  $E$ has the same value as in the non-resonant case, i.e., $E=-27/35$ \cite{Livertsetal}. The discussion concerning the values of the rest of the three coupling coefficients and their significance is deferred however until after the derivation of the solutions of eqs. (\ref{daughter11})-(\ref{parent11}).

\textit{Solution of the dynamical amplitude equations} -- Multiplying eqs. (\ref{daughter11}) and  (\ref{daughter21}) by $a_s^*$ and $a_2^*$, respectively, and summing the resulting equations yield:
\begin{equation}
\frac{d}{d\tau}[a_2a_s^*+a_2^*a_s]=0.
\end{equation}
Consequently, eq. (\ref{parent11}) may be written as the following Duffing equation with a constant forcing term:
\begin{equation}
\frac{d^2a_1}{d\tau ^2}= a_1 +E a_1^3 +\Gamma _{10},
\label{parent21}
\end{equation}
where $\Gamma_{10}=\Gamma _1 (a_{20}a_{s0}^*+a_{20}^*a_{s0})$, and $a_{j0}, j=2,s$ are the initial values of the corresponding amplitudes.  The value of $\Gamma_{10}$ varies within a wide range due to the arbitrariness of the initial data. The equation for $a_1$ may be solved now separately from those of the other two amplitudes. The value of $\Gamma _{10}$ determines the number of fixed points for $a_1$, whether it is one (for $|\Gamma _{10} | >2/\sqrt{-27E}$) or three (for $|\Gamma _{10}| <2/\sqrt{-27E}$). However, regardless of the value of $\Gamma _{10}$, the amplitude of the parent MRI, while initially growing exponentially, saturates and eventually oscillates nonlinearly in a bursty fashion with a constant amplitude, as is exemplified in Fig. (1). After solving for $a_1$, the dynamical equations for the daughter modes are easily solved by defining the following new time variable:
\begin{equation}
 \tau '= \sqrt{|\Gamma _2 \Gamma _s |}\int _0^{\tau} a_1(\xi)\,d\xi.
 \label{newvar}
\end{equation}
The nature of the solution of eqs. (\ref{daughter11}) and (\ref{daughter21}) depends now on $\sigma = sign(\Gamma _2 \Gamma _s)$, and is given by:
\begin{eqnarray}
a_2(  \tau ')&=& a_{20}\cosh(\sqrt{\sigma} \tau ')+i\sigma a_{s0}\alpha _2 \sinh(\sqrt{\sigma} \tau ') \label{solac}\\
a_s( \tau ')&=&a_{s0}\cosh(\sqrt{\sigma} \tau ')+i\sigma a_{20}\alpha _s\sinh(\sqrt{\sigma} \tau ') \label{solms},
\end{eqnarray}
where $\alpha_j=\Gamma_j/\sqrt{|\Gamma _2 \Gamma _s |}, j=2,s$.  When the daughter AC mode is a $k=2$ slow wave $\sigma$ can be shown to be equal to $1$.  Equations (\ref{solac}) and (\ref{solms}) reveal therefore the following result: If $\Gamma _2 \Gamma _s >0$ the linearly \textit{stable} AC and MS modes that participate in the resonant triad are nonlinearly destabilized by energy transfer from the linearly \textit{unstable} MRI mode, which is consequently saturated. An effective growth rate of the MRDI of the daughter modes may thus be estimated as
 $\gamma _{nl}=|\langle a_1\rangle |\sqrt{|\Gamma _2 \Gamma _s|}$, where 
$\langle a_1\rangle = \lim _{\tau \rightarrow \infty}\tau ^{-1} \int _0^{\tau} a_1(\xi)d\xi$.

\begin{figure}[h]
\centering
\includegraphics*[width=60mm,height=45mm]{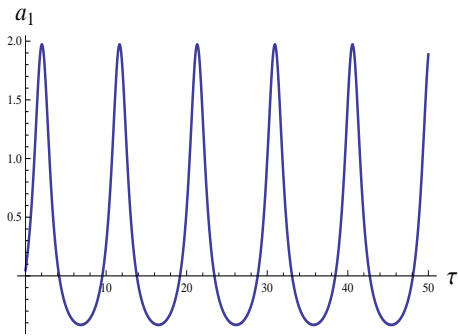}
\caption{Amplitude of the parent MRI. $a_1(0)=0.05, da_1/d\tau (0)=0.5, a_2(0)=1, a_s(0)=0, \Gamma _{10}=2/\sqrt{-27E}$}
\end{figure}
\begin{figure}[h]
\centering
\includegraphics*[width=60mm,height=45mm]{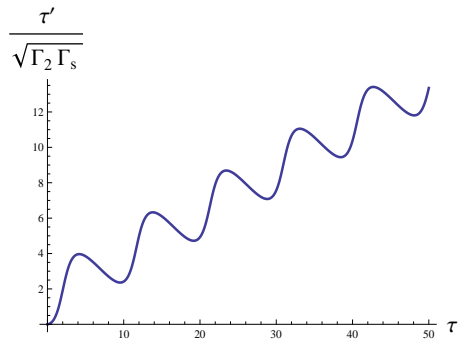}

\caption{$\tau '$ as a function of $\tau $ [see eqs. (\ref{newvar})-(\ref{solms})]. Same parameters as in Fig. 1.}
\end{figure}
\begin{figure}[h]
\centering
\includegraphics*[width=60mm,height=45mm]{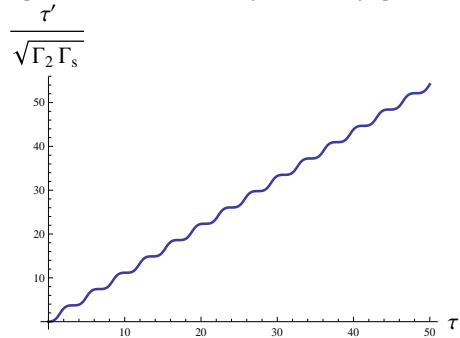}
\caption{$\tau '$ as a function of $\tau $ [see eqs. (\ref{newvar})-(\ref{solms})]. $a_1(0)=0.05, da_1/d\tau (0)=0.5, a_2(0)=1, a_s(0)=0, \Gamma _{10}=10/\sqrt{-27E}$.}  
\end{figure}

\textit{Results - } Figure (1) demonstrate the saturation of the MRI while figures (2) and (3) describe the simultaneous exponential growth of the daughter waves for two different values of $\Gamma _{10}$. It can be easily seen that the growth rate of the daughter waves does indeed depend on their initial conditions through the parameter $\Gamma _{10}$.  As the latter grows, so does $|\langle a_1 \rangle|$ and with it 
$\gamma _{nl}$. In addition, as $\Gamma _{10}$ grows, the steady-state solution for $a_1$ changes its nature from a three fixed-points solution to a single fixed-point one. This transition occurs for $|\Gamma _{10}|=2/\sqrt{-27E}$.

\textit{Conclusions - } The mechanism that is classically known as the decay instability is revisited and adapted to the geometry and physics of thin magnetized Keplerian disks. The resulting MDRI mechanism is conjectured to play an important role in the nonlinear evolution of the MRI. In the classical decay instability scenario developed for infinite homogeneous and immovable plasma, energy is transferred back and forth between a parent Alfv\'{e}n
wave and Alfv\'{e}n and acoustic daughter waves through a three-wave resonant interaction. The thin disk version of the decay instability that has been introduced in the current work is shown to deviate significantly from its classical predecessor. Instead of the classical stable Alfv\'{e}n wave, the role of the parent wave is currently played by an MRI mode that is slightly above the instability threshold. Hence, its amplitude is governed by a second order forced Duffing equation. The daughter waves are invariably AC and MS modes. In particular, it has been shown that for all possible initial conditions the parent MRI saturates in a bursty oscillatory manner. Furthermore, when the AC daughter wave is a slow AC mode, the linearly stable pair of daughter waves is nonlinearly destabilized and grow exponentially in time by tapping into the MRI energy. If, however, the role of the AC daughter wave is played by a stable fast mode, the amplitudes of all three modes remain bounded as they exchange energy periodically in a manner that resembles the classical decay instability. The picture of a resonantly interacting triad of modes may be easily generalized to a cluster of triads for a given parent MRI mode.

%
%
\textit{Acknowledgment -} This work was supported by grant number 180/10 of the Israel Science Foundation.

\end{document}